# Magnetic turnstiles in nonresonant stellarator divertor

Alkesh Punjabi[1,a)] and Allen H. Boozer[2,b)]
[1]Department of Mathematics, Hampton University, Hampton, Virginia 23668, USA
[2]Department of Applied Physics and Applied Mathematics, Columbia University, New York, New York 10027, USA
a)alkesh.punjabi@hamptonu.edu
b)ahb17@columbia.edu

**ABSTRACT:** Non-resonant stellarator divertors have magnetic flux tubes, called magnetic turnstiles, that cross cantori, which are fractal remnants of destroyed invariant tori with holes, that lie outside the outermost confining surface. The exiting and entering flux tubes can be adjacent as is generally expected but can also have the unexpected feature of entering or exiting at separate locations of the cantori. Not only can there be two types of turnstiles, but pseudo turnstiles can also exist. A pseudo turnstile is formed when a cantorus has a sufficiently large, although limited, radial excursion to strike a surrounding chamber wall. The existence of non-adjacent and adjacent turnstiles and pseudo turnstiles resolves issues that arose in earlier simulations of nonresonant stellarator divertors [A. Punjabi and A. H. Boozer, Phys. Plasmas **27**, 012503 (2020)].

**I. Introduction**

The fundamental properties of divertors in toroidal plasmas are determined by the magnetic flux tubes that carry the plasma from the plasma edge to the divertor chamber. Passage of field lines from outside the outermost surface to the wall can be characterized by the probability per toroidal transit that a magnetic field line will escape confinement and strike the surrounding wall. This probability increases as the radial separation from outermost confining magnetic surface becomes greater. The probability scales as some positive power of separation in $\psi_t$-space, where $\psi_t$ is toroidal magnetic flux. $\psi_t$ is defined by $\psi_t \equiv \pi B_c r^2$. $r$ is the radial position. $B_c$ is a characteristic magnetic field strength. This power is called the probability exponent [1,2]. For a sufficiently large perturbation to an axisymmetric divertor, the escaping flux tubes go from being determined by the singularity in the safety factor $q(\psi_t)$ on the separatrix to being determined by cantori/turnstiles as are stellarator divertors. Non-axisymmetric divertors are subject to more control than axisymmetric tokamak divertors.

The magnetic field lines outside the outermost confining magnetic surface in stellarator divertors travel to the wall through magnetic flux tubes. The magnetic field lines travelling from the wall towards outermost surface also enter the region outside the outermost surface through the magnetic flux tubes. These tubes always come in a pair. In one flux tube lines go out to the wall; and in the other tube, the lines from the wall come towards the outermost surface, preserving the magnetic flux. The outgoing and the incoming tubes together are called a magnetic turnstile. The magnetic turnstiles are like a pair of one-way bridges that connect the outermost surface to the wall. There can be more than one magnetic turnstile. The intersections of a magnetic turnstiles with the wall give the magnetic footprints on the wall. Calculation of the magnetic turnstiles is important for stellarator divertors. The calculation of magnetic turnstiles gives us the locations of the magnetic footprints on the wall, locations outside the outermost surface where the magnetic flux tubes exit and enter, the paths the outgoing and incoming lines take in the annulus between the outermost surface and the wall, the time it takes for the lines to traverse these paths, whether the footprints have fixed locations, how the footprints change as the wall is moved or if the geometry of wall is changed, how the footprints and paths change when different types of perturbations are present such as when the periodicity of the stellarator or the stellarator symmetry is broken. A host of questions on optimization of divertors can be addressed by calculation of magnetic turnstiles in divertors.

In this paper, a new simple and straightforward method to calculate the magnetic turnstiles in the nonaxisymmetric magnetic geometry of stellarator divertors is used. This method is applied to calculate the magnetic turnstiles in nonresonant stellarator divertor [1,2].

The calculation of the magnetic turnstiles in nonresonant stellarator divertor shows that there are two kinds of field line excursions outside the outermost confining surface. The first kind of excursions are true magnetic turnstiles; and the second kind of excursions are a pseudo turnstile. The pseudo turnstile is located outside the outermost surface in the layer of circulating field lines above the outermost surface where the surface has the largest radial excursion. We call this elevated layer a hump.

In the conventional double flux tube, the outgoing and incoming flux tubes exit and enter the outermost surface at adjacent locations. In the separated flux tube, the outgoing and the incoming flux tubes exit and enter outside the outermost surface at different locations. See Figures 2, 3, 5, and 7 below. In this paper, we will call the conventional double flux tube "adjoining turnstile", and the separated flux tube, as "separated turnstile". In the Figures, the adjoining turnstile will be indicated by the letter "A", and the separated turnstile by the letter "S". Both the adjoining and the separated turnstiles are true magnetic turnstiles.

In our model, the wall is an axisymmetric torus placed sufficiently far from the outermost confining surface. The radius of the wall is four times the minor radius. The method can also be used for walls with any geometry, axisymmetric or nonaxisymmetric. Optimization of the wall location can be easily included. This allows to one to test different types of walls for optimization. The outgoing and incoming tubes of the true turnstiles and the pseudo turnstile are stellarator symmetric [3]. This is consistent with the stellarator symmetry of the underlying magnetic configuration of stellarators. The probability exponents of the both the adjoining and separated turnstiles are calculated using the simulation method developed in [1,2]. The calculation of the turnstiles in this paper resolves the two questions about the magnetic turnstiles in the nonresonant stellarator divertor from our previous paper [2]: first, the nature of the continuous toroidal stripes in the footprints of the magnetic field lines; and second, the negative probability exponent. The answer for the continuous toroidal stripes is that they are not projection of the outermost confining surface on the wall; but are actually a true magnetic turnstile. It is the adjoining turnstile. The answer on the negative exponent is that the apparent footprint with negative exponent is the intersection of the pseudo turnstile with the wall. The footprint is formed when the lines from the layer of circulating lines above the hump on the outermost surface reach the farthest wall. This happens only when the lines are given a sufficiently large radial displacement. See Section IIID below. The hump is located above the outmost surface at the position of the largest radial excursion of the outermost surface. It is not so much the radial displacement but whether the magnitude of the distortions in the Cantori outside the last confining surface are sufficiently large to reach a nearby wall even though the lines would not reach an arbitrarily distant wall as true turnstiles would.

Some of the invariant tori in unperturbed Hamiltonian systems are destroyed when a perturbation is present. When an invariant torus with the rotation frequency $f$ is destroyed at a critical value of the perturbation amplitude, and if $\frac{f}{\omega}$ is a rational number, then a chain of resonances or islands of stability appears in its place. $\omega$ is the frequency of the perturbation. If $\frac{f}{\omega}$ is an irrational number, then a cantorus appears in the place of that invariant torus. Cantorus is a Cantor invariant set. The motion on a cantorus is unstable and quasiperiodic.

Because the motion on a cantorus is unstable, it has stable and unstable manifolds. Each orbit on a cantorus is dense, but just like a torus, there are uncountably many orbits. Cantori do not occupy a volume of a finite measure in the phase space. Cantori form an infinite hierarchy around islands of stability. The closer a cantorus is to the island's boundary, the narrower are its gaps.

In the phase space, cantori essentially influence the transport. This is because it may take long time for particles and their trajectories to percolate through the cantori gaps. The smallest gaps in cantori appear when cantori are near the boundaries of islands of stable trajectories. In this region, the density of points in the Poincaré section is high. The regions just outside the islands are called dynamical traps. The long stay in dynamical traps is called stickiness. Cantori are not the only reason for stickiness and dynamical traps. Hyperbolic trajectories along with their stable and unstable manifolds also produce complicated tangles where particles and their trajectories can be trapped for a long time.

Cantori are intrinsically defined in area-preserving twist maps and their corresponding Hamiltonian systems. On the other hand, turnstiles are not intrinsically defined. The term turnstile was first coined in 1984 by MacKay *et al* [4,5]. In [4,5], the turnstiles of a cantorus were constructed. This was achieved by first choosing one gap in the cantorus and closing its forward images by pieces of stable manifold and backward images by pieces of unstable manifold and taking both in the chosen gap. The choice of the gap was arbitrary. Further, the escaping and entering fluxes can also be chosen to be distributed in many different ways between the gaps. Choices of gaps and the choices of fluxes that pass through the gaps can give different turnstiles, including adjoining and separated turnstiles. It is a question of how one chooses to close the gaps in cantori. This topic is reviewed by MacKay in [6]. Another review of this topic is by Meiss [7].

Our application of turnstiles to nonresonant stellarator divertor here is different. Our goal is to calculate the magnetic turnstiles or flux tubes that take the magnetic field lines from outside the outermost surface to the wall and from the wall to outside the outermost surface. We take points on the outermost surface in the $\zeta = 0$ poloidal plane and give them a random radial outward kick with the maximum kick of about 1% of the minor radius. Field lines start their trajectories at these shifted positions. These field lines make their journeys to the wall percolating through cantori, tangles, chains of small islands, and chaos in the annuli between outermost surface and the wall. They encounter dynamical traps and tangles. Our purpose is to find out if the lines form magnetic flux tubes under the accumulative effects of traps and tangles. Our application of turnstiles here is practical in nature and it has physical relevance for the stellarator divertors. This exercise provides an important perspective on the behavior/trajectories of magnetic field lines near the edge of the plasma through to intersection with the wall. This is an important topic for understanding 3D fusion plasmas generally. The complexities and possibilities for divertors that may arise in the stellarator divertors are particularly interesting. We have identified an unexpected property of magnetic field line flow, that the entering and exiting flux tubes are not always adjacent.

The overarching motivation for this paper is to develop a general method to create a 3D picture of magnetic turnstiles in stellarator divertors. In the method, the phase space is divided into large number of cells. The cells cover the entire volume between the magnetic axis and the wall. The lines are started outside the outermost confining surface. The lines are advanced forwards and backwards for a sufficiently long time. After each step of integration, the position of line is calculated. From this position, we determine the cell which the field line is visiting. For every visit of a given cell, the population count of the cell is elevated by unity. The integration of a field line is stopped when the line reaches the farthest wall at $r = r_{WALL} = 4b$ where $b$ is the minor radius of the stellarator. The 3D distribution of the nonempty cells gives us the magnetic turnstiles. We have used this method to calculate the magnetic turnstiles in the nonresonant stellarator divertor. The magnetic configuration that we have used here is the same as that we used in our simulation of nonresonant stellarator divertor [2].

This paper is organized as follows: In Section II, we apply this method to calculate the magnetic turnstiles in the nonresonant stellarator divertor. In Section III, we give the results. In Section IV, we give the summary and conclusions, and discuss the results.

## II. Calculation of magnetic turnstiles in nonresonant stellarator divertor

### A. Magnetic configuration

The magnetic configuration of nonresonant stellarator divertor used for this study is same as in [2]. The Hamiltonian for the trajectories of magnetic field lines in nonresonant stellarator divertor is given by [2]

$$\frac{\psi_p}{\bar{\psi}_g} = \left[\iota_0 + \frac{\varepsilon_0}{4}((2\iota_0-1)\cos(2\theta-\zeta)+2\iota_0\cos2\theta)\right]\left(\frac{\psi_t}{\bar{\psi}_g}\right)$$

$$+\frac{\varepsilon_t}{6}[(3\iota_0-1)\cos(3\theta-\zeta)-3\iota_0\cos3\theta]\left(\frac{\psi_t}{\bar{\psi}_g}\right)^{3/2}$$

$$+\frac{\varepsilon_x}{8}[(4\iota_0-1)\cos(4\theta-\zeta)+4\iota_0\sin4\theta]\left(\frac{\psi_t}{\bar{\psi}_g}\right)^2.$$

$\psi_p$ is the poloidal flux, $\psi_t$ is the toroidal flux, $\iota_0$ is the rotational transform on magnetic axis, $\zeta$ is the toroidal angle of the single period. The poloidal flux $\psi_p(\psi_t, \theta, \zeta)$ is the Hamiltonian function for the trajectories of magnetic field lines. The toroidal flux $\psi_t$ and the poloidal angle $\theta$ are canonically conjugate for evolution in the toroidal angle $\zeta$. $\zeta$ is related to the toroidal angle $\varphi$ by $\zeta = n_p\varphi$. $\theta$ is the poloidal angle. The radial position $r$ is given by $r \equiv \sqrt{\psi_t/\pi B_c}$. $B_c$ is a characteristic magnetic field strength. The poloidal flux $\psi_p$ is per period. The rotational transform per period is $\iota_0 \cong 0.15$. The shape parameters $\varepsilon_0, \varepsilon_t,$ and $\varepsilon_x$ control the elongation, triangularity, and the sharpness of the edges of the outermost confining surface, respectively. Here $\varepsilon_0 = \varepsilon_t = 1/2$ and $\varepsilon_x = -0.31$. $\bar{\psi}_g$ is an averaged toroidal flux used to normalize magnetic fluxes. The stellarator has five periods, $n_p = 5$; and all the five periods are identical. The step-size of the map is $\delta\zeta = 2\pi/3600$, same as used in [2]. The phase portraits are shown in Figure 1 in [2]. The outermost confining surface is at $r/b \cong 0.87$, $\theta = 0$, and $\zeta = 0$ [2]. The average normalized toroidal flux inside the outermost confining surface is $\psi_0 = 1.3429$. The wall is circular with radius $r_{WALL}/b = 4$. $r_{WALL}$ is the radius of the wall and $b$ is the minor radius. The furthest the outermost confining surface extends in radial direction is $r_{MAX}/b = 1.9998$ [2] and the smallest extension is $r_{MIN}/b = 0.73$. See Figure 1 in [2]. The rotational transform $\iota(r)$ for a single period is same as in [2]. See Figure 2 in [2].

**B. Choice of starting positions of field lines**

The outermost confining surface is at $r/b = 0.87$, $\zeta = 0$, $\theta = 0$. We start a field line at this point and advance it for 10,000 toroidal circuits of a period. Every time this field line crosses the $\zeta = 0$ poloidal plane, we record its puncture point $(\theta_j, \psi_{t,j})$, $j = 1,2,...,10000$. We separate out every 10$^{th}$ point from these 10,000 points, i.e., $j = $ 1,10,20,30,...,10000. These are 1000 points in the $\zeta = 0$ plane on the outermost surface. We use the subscript OCS to denote the outermost confining surface. The radial positions of these 1000 points, $r_{OCS}^{(j)}$, $j = 1,2,...,1000$, are then shifted radially outwards along the line from the O-point keeping the poloidal positions $\theta^{(j)}$ unchanged. The outward shift is random and the maximum shift is $\Delta r = 10^{-2}$; $r_0^{(j)} = r_{OCS}^{(j)} + R^{(j)}\Delta r$. $r_{OCS}^{(j)}$ is the radial position of the $j^{th}$ point on the outermost surface, $R^{(j)}$ is a random number in the interval $[0,1)$. With $\psi_{t,0}^{(j)} = \left(r_0^{(j)}\right)^2$, this gives us the initial conditions $\left(\theta_0^{(j)}, \psi_{t,0}^{(j)}\right)$, $j = 1,2,...,1000$, in the $\zeta_0 = 0$ plane. We show the starting positions of the field lines in Figure 1.

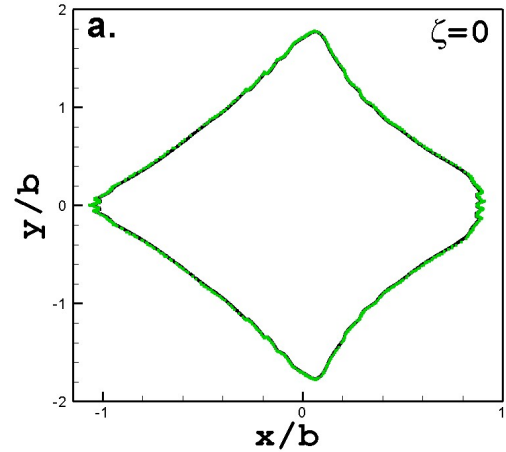

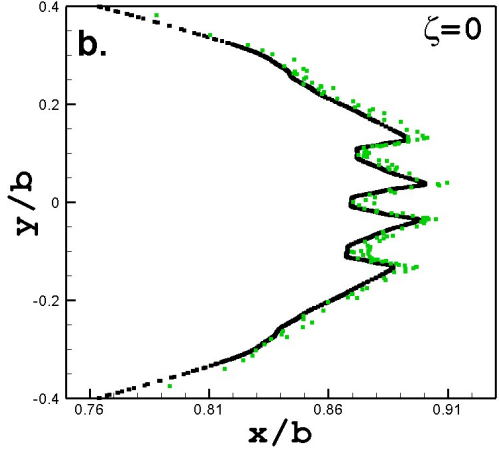

FIGURE 1. (a) The starting positions are calculated by random radial shifting of the points on the outermost surface. Maximum shift is $\Delta r = 10^{-2}$. The outermost surface and the starting positions in the $\zeta = 0$ plane are shown, (b) An enlarged view of Figure 1(a). Color code: Black = the outermost surface, Green = the starting positions.

**C. Calculation of magnetic turnstiles**

These 1000 initial points represent the field lines very close to and outside the outermost surface. These lines are advanced forward and backward for 200,000 toroidal circuit of the period using the stellarator map in [2]. The stellarator map is an area-preserving map.. The map equations are given in Equations (2)-(4) in [2]. The step-size of the map is as before, $\delta\zeta = 2\pi/3600$.

To calculate the magnetic turnstiles inside the axisymmetric torus of radius $r_{\text{WALL}}/b = 4$, the torus is divided into 360×360×400 cells. These 3D cells are each of the size $\Delta\zeta \times \Delta\theta \times \Delta r = (2\pi/360) \times (2\pi/360) \times (b/100)$. A 3D integer array $C(360,360,400)$ is used. The array is initialized to $C(i,j,k) \equiv 0$ for all $i$, $j$, and $k$. For a given line, after each iteration of the map, the position of field line $\left(\zeta_l^{(n)}, \theta_l^{(n)}, r_l^{(n)}\right)$ is calculated. The subscript $l$ denotes the field line and the superscript $n$ denotes the iteration. If $(i-1)\Delta\zeta \leq \mathrm{mod}\left(\zeta_l^{(n)}, 2\pi\right) < i\Delta\zeta$, $(j-1)\Delta\theta \leq \mathrm{mod}\left(\theta_l^{(n)}, 2\pi\right) < j\Delta\theta$, and $(k-1)\Delta r \leq r_l^{(n)} < k\Delta r$, the value of $C(i,j,k)$ is increased by unity from its previous value. This is done after each iteration of all the lines that strike the wall before or at the end of 200,000 toroidal transits of the period. If the line strikes the wall before the end of 200,000 transits, the integration of the line is terminated after it strikes the wall. The array $C$ does not include the occupancy data for lines that do not reach the wall before or at the end of 200,000 circuits. The same procedure is used for the 1000 lines when they move backwards. A similar procedure is used to calculate the 3D picture of the outermost confining surface. The $C$ arrays for the forward lines, $C_{\text{FW}}$; the backward lines, $C_{\text{BW}}$; and the outermost surface, $C_{\text{OCS}}$, give us the three-dimensional picture of the magnetic turnstiles in the annulus between the outermost surface and the wall in the divertor.

$C(i,j,k) = 0$ tells us that no field line ever goes through those cells. High counts of $C(i,j,k)$ tells us that field lines go to those cells often, meaning that those regions are sticky.

**III. Results**

From the forward moving 1000 lines, 645 lines; and from the backward moving lines, 657 lines reach the wall at $r_{\text{WALL}}/b = 4$.

**A. Magnetic turnstiles**

The magnetic turnstiles in poloidal planes $\zeta = 0$, $\pi/2$, $\pi$, and $3\pi/2$ are shown in Figures 2 and 3. The Figure 2 shows turnstiles in $(x,y)$ coordinates and the Figure 3 shows them in the ($\theta$,r) coordinates. The adjoining turnstile is identified by the label A, and the separated turnstile is identified by the label S. The adjoining turnstile is located slightly below $\theta = 2\pi$ for outgoing trajectories, and slightly above $\theta = 0$ for the incoming trajectories. The adjoining turnstile leaves and enters outside the outermost surface at the same location. That is why we call it the adjoining turnstile. The adjoining turnstile covers the entire range of the toroidal angle of the period from $\zeta = 0$ to $2\pi$. See Figure 4. The separated turnstile does not cover the whole range of the toroidal angle. The lines in the separated turnstile reach the wall in a restricted range of toroidal angle. See Figures 2 and 3.

The unexpected result is that the separated turnstile does not enter and leave the region outside the outermost surface at adjacent locations; the outgoing and the incoming tubes of the separated turnstile leave and enter at different locations outside the outermost surface. See Figures 2, 3, and 5.

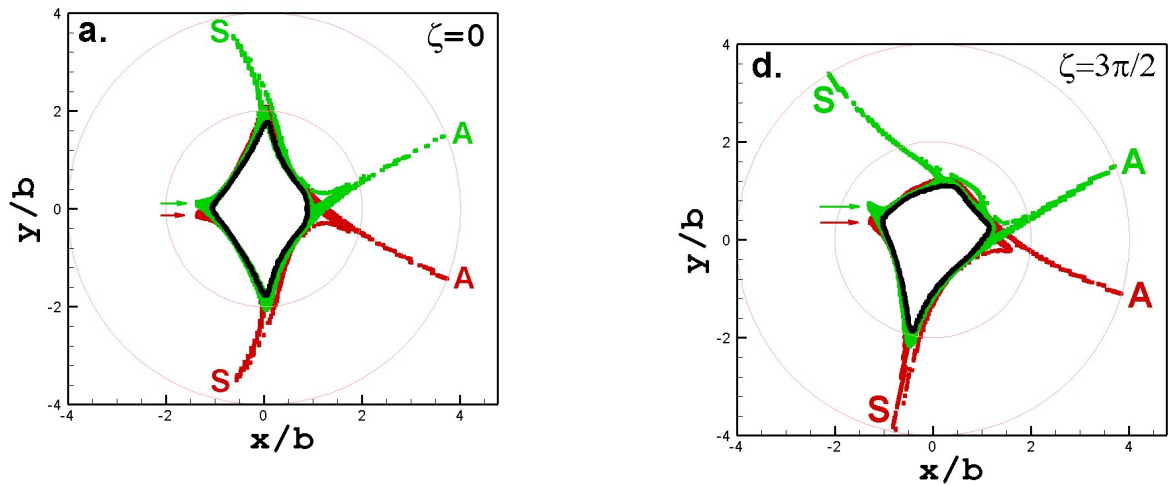

FIGURE 2. The magnetic turnstiles in (x,y) coordinates in poloidal planes (a) $\zeta = 0$, (b) $\zeta = \pi/2$, (c) $\zeta = \pi$, and (d) $\zeta = 3\pi/2$. Color code: Black = the outermost surface, Red = forward trajectories, Green = backward trajectories,

Grey = axisymmetric tori $r/b = 2$ and $r/b = 4$. The largest radial reach of the outermost confining surface is $r/b = 1.9998$. So, $r/b = 2$ is the first wall that does not intersect the outermost surface. $r/b = 4$ is the wall. The labels A and S denote the adjoining turnstile and the separated turnstile, respectively.

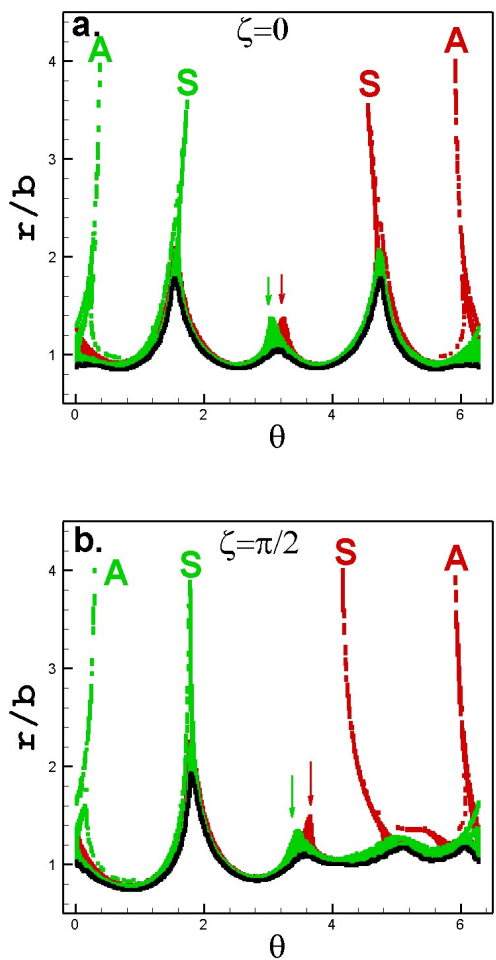

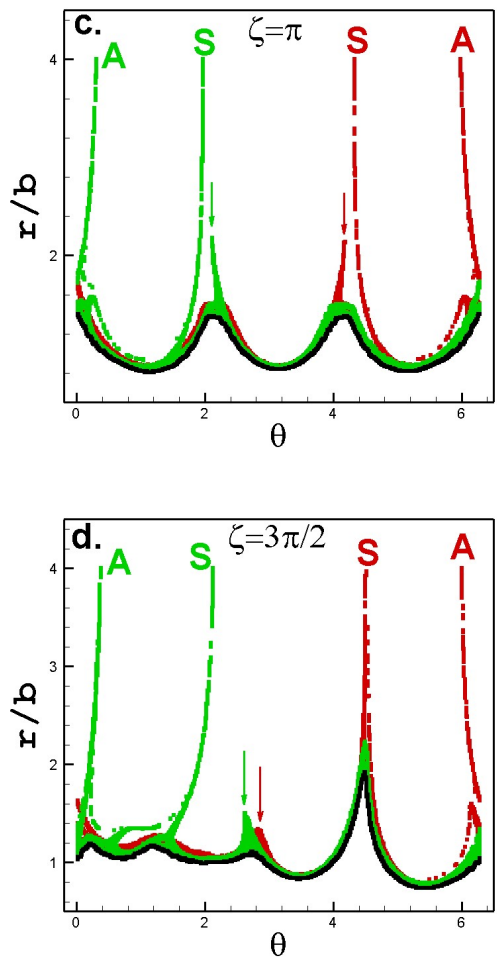

FIGURE 3. The magnetic turnstiles in $(\theta, r)$ coordinates in the poloidal planes (a) $\zeta = 0$, (b) $\zeta = \pi/2$, (c) $\zeta = \pi$, and (d) $\zeta = 3\pi/2$. Color code: Black = the outermost surface, Red = forward trajectories, Green = backward trajectories.

**B. Footprints**

From the $C$ arrays for the forward and backward lines, we can calculate the footprints on walls of radii $r/b$ separated by $\Delta r/b = 0.01$ from $r/b = 0.01$ to 4. The largest radial reach of the outermost surface is $r_{MAX}/b = 1.9998$. A circular first wall must have a radius $r/b > 1.9998$ not to intercept the outermost confining surface. The furthest wall is at $r = 4b$. So, the first wall that does not intersect the outermost surface is $r/b = 2$.

The final wall is at $r = r_{WALL} = 4b$. $b$ is the minor radius.

At $r/b = 2$, all lines go only into three stripes on the wall. This continues from $r/b = 2$ to $2.38$. At $r/b = 2.39$, one stripe disappears and all the lines go only into two stripes. These two stripes are the intersections of the adjoining turnstile and the separated turnstile with the walls. This continues from $r/b = 2.39$ to $4$. At $r/b = 4$, we terminate the field lines. For the forward footprints, the true turnstiles can be identified by their locations on the walls. The outgoing adjoining turnstile is confined to $\theta \geq 5.5$. The adjoining turnstile covers the full range of the toroidal angle, $0 \leq \zeta < 2\pi$. The outgoing adjoining turnstile is confined to $\pi \leq \theta < 5.5$. The incoming adjoining turnstile is confined to $\theta < 1$; and the incoming separated turnstile to $1 \leq \theta < \pi$. See Figure 4.

The locations of the footprints on the walls are fixed. This is consistent with the findings of Bader *et al* [8]. Bader *et al* paper [8] uses fields calculated using the HSX coils. The footprint becomes smaller in size as the wall moves out. The forward and backward footprints are stellarator-symmetric [3]. See Figure 4.

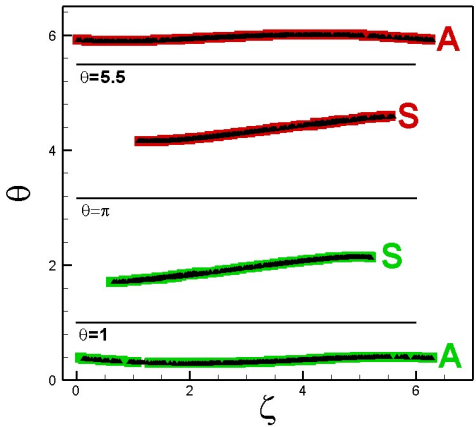

Figure 4. The footprints on the wall at $r/b = 4$. The footprints are where the turnstiles cross the wall. The labels A and S denote the adjoining and separated true magnetic turnstiles, respectively. The footprints from the simulation [2] for the smallest velocity $u_\psi = 2 \times 10^{-5}$ is also shown as black unfilled triangles. The footprints have fixed locations on the walls and they are stellarator symmetric. Color codes: Red = the forward footprints, and Green = the backward footprints, Black = footprints from simulation for velocity $u_\psi = 2 \times 10^{-5}$.

Some of the 645 forward lines and some of the 657 backward lines go into the hump (see Section I. Introduction above) for some time when $r/b \leq 2.38$. These humps are indicated by a red and a green arrow for the forward and the backward trajectories, respectively, in Figures 2 and 3.

**C. Calculation of true magnetic turnstiles**

645 outgoing and 657 incoming lines reach the wall through true magnetic turnstiles. Out of 645 forward lines, 390 go into the adjoining turnstile and 255 go into the separated turnstile. Out of 657 backward lines, 391 go into the adjoining turnstile and 266 go into the separated turnstile. For $r/b \geq 2$, the tori with radius $r/b > 2$ do not intersect the outermost surface. We separate out the 390 forward lines that go into the adjoining turnstile. We advance them for 200,000 circuits. We repeat the same for the 255 forward lines going into the separated turnstile. We do the same for the 391 and 266 lines of the backward moving lines. From this, we calculate the adjoining and the separated true magnetic turnstiles from the forward and backward moving lines. We show the true turnstiles in Figure 5.

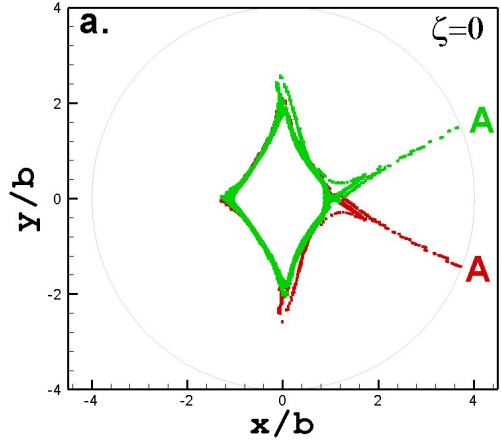

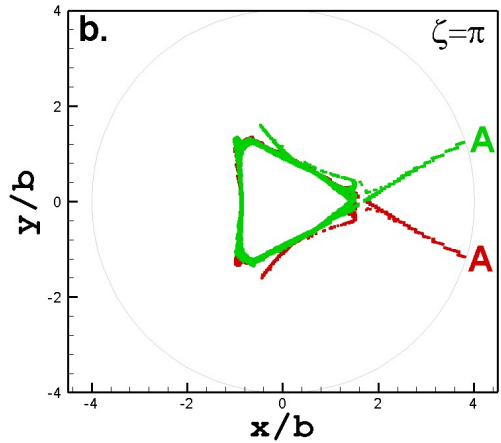

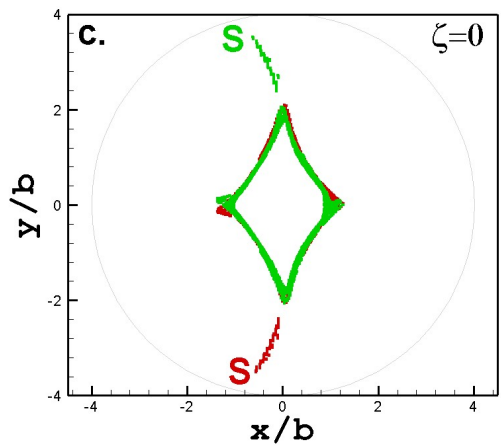

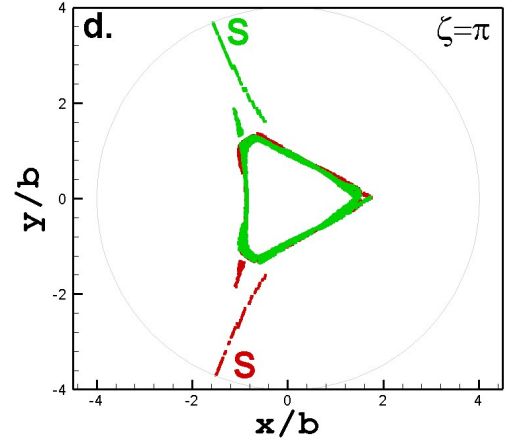

FIGURE 5. The true magnetic turnstiles in the nonresonant stellarator divertor. The outgoing and incoming turnstiles of the adjoining turnstile in the plane (a) $\zeta = 0$ and (b) $\zeta = \pi$; and the separated turnstile in the plane (c) $\zeta = 0$ and (d) $\zeta = \pi$.

## D. Answers to open questions from the simulation

In our simulation of nonresonant stellarator divertor [2], for all values of velocities, some field lines formed a pair of continuous toroidal stripes on the wall. The nature of these continuous toroidal stripes was not understood, and we had said that it needed further study [2]. This was the first open question from the simulation.

We had also found that there were two kinds of magnetic turnstiles. We had called them the primary and the secondary turnstiles. The exponent for the primary turnstile was 9/4 and the exponent for the secondary turnstile was -3/2. The secondary turnstile existed only for sufficiently large values of the radial velocity. The existence of the secondary turnstile only for sufficiently large speeds and its negative exponent were unexpected [2]. This was the second open question from the simulation.

Here we show that the calculation of

turnstiles in this paper answers these questions.

We first compare the footprints on the final wall $r/b = 4$ from the magnetic turnstiles here with the footprints from the lowest velocity $u_\psi = 2\times 10^{-5}$ in the simulation [2]. These are shown in Figure 4. We see that the two footprints match as expected.

The footprints of adjoining turnstile coincide with the continuous toroidal stripes covering the full range of the toroidal angle of the single period found in the simulation [2]. This answers the first question. The nature of the toroidal stripes is that of a true magnetic turnstile, and it is the adjoining turnstile found here.

Some of the 645 forward lines and some of the 657 backward lines that reach the final wall, go into a hump for some time when $r/b \leq 2.38$. These humps were called the secondary family in [2]. There we had found that the probability exponent for this pseudo turnstile was $d = -3/2$. There we had said that the probability exponents are expected to be positive and that the negative exponent is unexpected.

The hump on the forward moving trajectories is located in the poloidal range 1.5 to 1.92. The hump is a protruding layer of circulating field lines. It has a sharp edge. The largest radial protrusion of the hump is $r/b = 2.38$ at $\theta = 1.71$. The hump is located above the outermost surface where the surface has largest radial protrusion. The distinction between the hump reaching $2.38b$ while the outermost surface reaching $1.9998b$ is that the region between the two is a layer of circulating field lines. The sharp edge of the hump at $\theta = 1.7104$ very closely coincides with the location of the largest radial excursion of the outermost confining surface at $r_{\text{MAX}}/b = 1.9998$, $\zeta = 0.9949$, and $\theta = 1.6986$ [2]. If the field lines were given sufficiently large velocity, the lines from the layer of circulating lines in this hump could reach the final wall, giving the footprint of the secondary turnstile with a negative probability exponent seen in the simulation [2] when the radial velocity is sufficiently large. See Section IIID below. We identify the secondary turnstile of simulation with the pseudo turnstile found here. This also explains why there was a terminal velocity for the pseudo turnstile. See Section IIID below. This answers the second question from the simulation.

What we had called the continuous toroidal stripes in the simulation is what we call the adjoining turnstile here; and what we called the primary family in [2] is what we call the separated turnstile here; and what we called the secondary family then is what we call the pseudo turnstile here.

**E. Calculation of probability exponent of the adjoining turnstile**

The probability exponent of a magnetic turnstiles is an important parameter. It tells us how fast the field lines reach the wall through the turnstiles. The probability exponent characterizes the probability of escape of field lines as a function of radial distance from the outermost surface for the field lines in a magnetic turnstile. The probability per radian advance in toroidal angle $\zeta$ that a magnetic field line will be lost by passing through a turnstile, $P_t(\psi_t)$, is modelled by a power-law when $\psi_t > \psi_0$. Then the probability is given by $P_t(\psi_t) = (d+1)c_p\left(\frac{\psi_t - \psi_0}{\psi_0}\right)^d$. See equation (22) in [1]. $d$ is called the probability exponent of the magnetic turnstile. The probability scales as the power $d$ of the separation from the outermost surface. We used the exponent $d$ to characterize loss of lines through magnetic turnstiles in [1,2]. We used it to fit the simulation results in [1,2]. $c_p$

is a normalizing constant. It was shown in [1] that the field lines loss-time through a turnstile is given by $\xi_l = \frac{1}{c_p}\left(\frac{c_p}{u_\psi}\right)^{\frac{d}{d+1}}$. See Equation (31) in [1]. The loss-time through a turnstile is the time it takes for the fraction of field lines remaining in the plasma to drop by one e-fold. The loss-time is counted from when the first field line strikes the wall. $u_\psi$ is the radial velocity given to field lines in $\psi_t$-space in the simulation [1,2]. In the simulation [1,2], field lines are started on a good surface located midway between the magnetic axis and the outermost surface. The field lines starting on this midway surface are given a constant radial velocity $u_\psi$ per radian of toroidal advance. Velocity is given in $\psi_t$ space. The constant values of velocity in [1,2] were 2E-5, 3E-5, 4E-5,…,1E-2 [2]. Simulation gives the loss time as a function of the velocity, $\xi_l(u_\psi)$. Linear fit to log($\zeta_l$) versus log($u_\psi$) gives $\zeta_l = c u_\psi^p$. Then, $p = -\frac{d}{d+1}$, or $d = -\frac{p}{p+1}$.

Both the underlying magnetic equilibrium and the wall are stellarator symmetric [3]. So, ideally the loss-times of the forward and the backward lines for a given velocity must be exactly equal. Departures from the exact equality occur due to statistical and numerical errors in the simulation. When the departures are sufficiently small, the simulation data can give us reliable estimates of scalings and the probability exponents. The deviation of the loss-times from exact equality of the loss-times of the forward and the backward lines for a given velocity is $\delta \xi_l = \Delta \xi_l / \langle \xi_l \rangle$ where the deviation is $\Delta \xi_l = |\xi_{l,FW} - \xi_{l,BW}|$ and the average loss-time is $\langle \xi_l \rangle = (\xi_{l,FW} - \xi_{l,BW})/2$ for a given $u_\psi$. The subscripts FW and BW refer to the forward and the backward field lines for a given velocity $u_\psi$.

We went back to our data in the simulation [2] for the toroidal stripes. We used the data when the deviation of the forward and backward loss times from the average is less than 2%, i.e., $\delta \zeta_l < 2\%$. With this criterion, we get 12 data points. We used these 12 data points to estimate the scaling of the loss-time with velocity for the adjoining turnstile, see Figure 6. Linear fit to log($\zeta_l$) versus log($u_\psi$) gives

$$\zeta_l = c u_\psi^p.$$

where $c = 1.3163$, $1.1222 < c < 1.5439$, $p = -0.6417 \pm 0.0198$, and the coefficient of multiple determination for the fit is $R^2 = 0.9906$. The probability exponent for the adjoining turnstile is $d_A = 1.7911$ and the error in $d_A$ is $1.6450 < d_A < 1.9543$. From this, $d_A = 1.7911 \cong 1.8 = 1\frac{4}{5} = \frac{9}{5}$ for the adjoining turnstile. So,

$d_A = 9/5$   adjoining turnstile.

From the simulation, we know the probability exponent of the separated turnstile. It is:

$d_S = 9/4$   separated turnstile.

The probability exponents for the adjoining and the separated turnstiles are not equal. So, the scalings are not universal [9]. There are two true magnetic turnstiles, the adjoining turnstile and the separated turnstile, with probability exponents of 9/5 and 9/4, respectively.

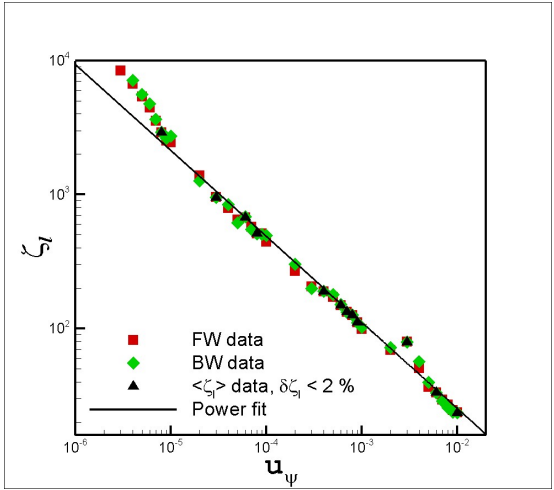

FIGURE 6. The scaling of the loss-time $\zeta_l$ with the velocity $u_\psi$ for the adjoining turnstile. Color code: Red squares = the loss time for the forward lines, Green diamonds = the loss-time for the backward lines, Black triangles = the average loss-time $\langle\zeta_l\rangle$ when the deviation of loss-times from exact equality $\delta\zeta_l < 2\%$, Black line = the power fit to $\langle\zeta_l\rangle$. The loss-time $\zeta_l$ scales as $1/u_\psi^{0.6417}$, giving the probability exponent $d_A = 1.7911$.

## F. Magnetic turnstiles of the nonresonant stellarator divertor when stellarator symmetry is broken

We add a stellarator symmetry breaking term in the poloidal flux given in Section IIA. This is done by replacing the shape parameter $\varepsilon_0$ by $\varepsilon_0(1+\delta\sin(\zeta))$. $\zeta$ is toroidal angle of the period, and $\delta$ is the amplitude of the stellarator symmetry breaking perturbation. With broken stellarator symmetry, the outermost confining surface shifts inwards to $r/b \cong 0.86$ when $\delta = 10^{-2}$. We calculate the 1000 new starting positions using the procedure described in Section IIB above. The maximum shift is $\Delta r = 10^{-2}$ as before. The perturbed magnetic turnstiles are shown in Figure 7.

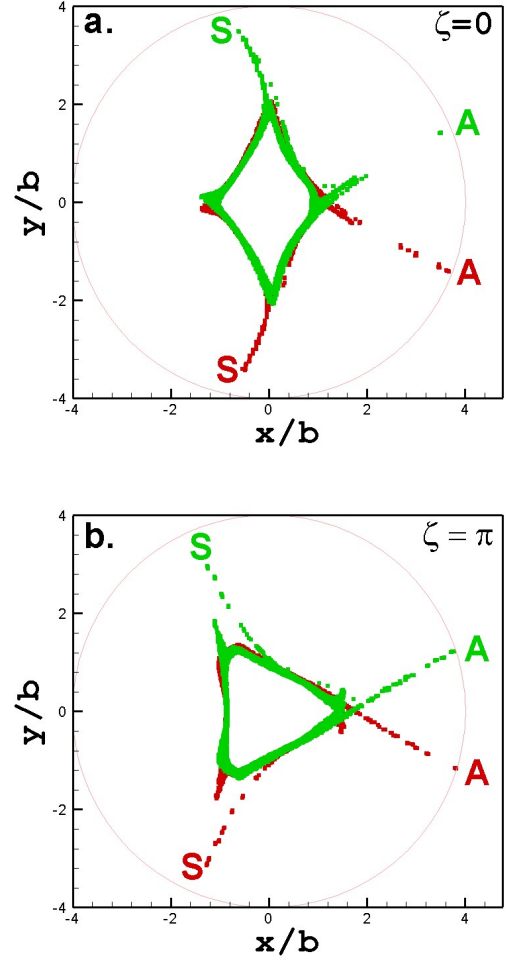

FIGURE 7. Magnetic turnstiles of the nonresonant stellarator divertor when stellarator symmetry is broken. (a) In the poloidal plane $\zeta = 0$, and (b) $\zeta = \pi$. Color code: Red = outgoing turnstiles, Green = incoming turnstiles, Grey = the wall at $r/b = 4$.

From Figure 7, we see that the locations of the magnetic turnstiles are robust against strong stellarator symmetry breaking perturbation. Even when the symmetry is broken, the separated magnetic turnstile leaves and enters at different locations outside the outermost surface.

## IV. Summary, conclusions, and discussion

We have calculated the magnetic turnstiles in the full 3D nonaxisymmetric magnetic geometry of nonresonant stellarator divertor. We have found that there are two true magnetic turnstiles – the adjoining turnstile and the separated turnstile - with probability exponents 9/5 and 9/4. The outgoing and the incoming magnetic turnstiles are stellarator symmetric. The footprints on walls have fixed locations. Our study has also answered the two questions from our simulation [2] of nonresonant divertor. The continuous toroidal stripes are a distinct type of true turnstiles; and what we called the secondary family of turnstile in our simulation is a pseudo turnstile. It is formed by lines that enter and leave a layer of circulating lines outside the outermost surface the outermost surface has largest radial excursion.

The unexpected result of our study is that the magnetic flux tubes of separated turnstile do not enter and leave at adjacent locations; the outgoing and the incoming tubes of the separated turnstile leave and enter at different locations outside the outermost surface.

These two turnstiles, with the separated turnstile leaving and entering at different locations, and the adjoining turnstile entering and leaving at adjacent locations, is robust when the stellarator symmetry is strongly broken.

The questions such as: How sensitive are these structures to small variations in the plasma currents? Such variations include the small fluctuations in the plasma currents which may be ultimately uncontrollable. Can they significantly change the geometry of the turnstiles? This would be important information, and in that case how should we proceed? These are the questions of the resiliency of the structures. The resiliency of the magnetic turnstiles can be tested by examining the changes in strike point patterns in the magnetic footprints which arise from such variations [8]. Bader et al have described this test in [8]. In this paper, they describe an initial description of the resilient divertor properties of quasi-symmetric (QS) stellarators using the HSX (Helically Symmetric eXperiment) configuration as a test-case. Divertors in high-performance QS stellarators will need to be resilient to changes in plasma configuration that arise due to evolution of plasma pressure profiles and bootstrap currents for divertor design. They tested the resiliency by examining the changes in strike point patterns from the field line following which arise due to configurational changes. In this test, a low strike point variation with high configuration changes corresponds to high resiliency. The HSX edge displayed resilient properties with configuration changes arising from the wall position, plasma current, and external coils. The resilient behavior was lost if large edge islands intersected the wall structure. The resilient edge properties were corroborated by heat flux calculations from the fully 3-D plasma simulations using EMC3-EIRENE. Further, the strike point patterns were found to correspond to high curvature regions of magnetic flux surfaces.

For example, if the Hamiltonian function for the trajectories of field lines in the W7-X stellarator is given, using the method described here, the magnetic turnstiles in W7-X divertor can be calculated. If the perturbed W7-X field line Hamiltonian is specified, the resiliency of the turnstiles in W7-X can be tested using the method given in [8].

There is more to plasma transport than free-streaming along field lines. How important are the turnstiles when there are collisions, electric fields, and so on? These effects are complicated, and are of practical consequences in the edge physics of stellarator divertors. These effects will require further investigations in future.

**Acknowledgements**

This material is based upon work supported by the U.S. Department of Energy, Office of Science, Office of Fusion Energy Sciences under Awards DE-SC0020107 to Hampton University and DEFG02-03ER54696 to Columbia University. This research used resources of the NERSC, supported by the Office of Science, US DOE, under Contract No. DE-AC02-05CH11231. We thank the reviewers for valuable comments and suggestions which helped us improve the paper and its clarity.

**Data availability**

The data that support the findings of this study are available from the corresponding author upon reasonable request.

**Conflict of interest**

The authors have no conflicts to disclose.